\documentstyle[12pt]{article}
\begin{document}
\begin{titlepage}
\begin{flushright}
                Preprint IFT UWr 922/99\\
                January 1999 \\
\end{flushright}
\bigskip
\bigskip

\begin{center}
{\Large \bf
Spectrum Generating Algebra\\
and\\
No-Ghost Theorem\\
for\\
Fermionic Massive  String\\}
\end{center}
\bigskip
\bigskip

\begin{center}
{\bf Zbigniew Hasiewicz\footnote{
                Institute of Mathematics, University in Bia{\l}ystok,
                ul.Akademicka 2, 15-267 Bia{\l}ystok, Poland;\\
                E-mail: zhas@alpha.fuwb.edu.pl},}\bigskip\\
        Institute of Mathematics, University in Bia{\l}ystok,
        Bia{\l}ystok, Poland \bigskip\\
{\bf Zbigniew Jask\'{o}lski\footnote{
                Institute of Theoretical Physics, University of Wroc{\l}aw,
                pl. Maxa Borna 9, 50-204 Wroc{\l}aw, Poland; \\
                E-mail: jaskolsk@ift.uni.wroc.pl}, and
Andrzej Ostrowski}\bigskip\\
        Institute of Theoretical Physics, University of Wroc{\l}aw,
        Wroc{\l}aw, Poland\\
\end{center}
\bigskip \bigskip
\bigskip\bigskip

\begin{abstract}
The covariant operator quantization of  the
ordinary free spinning BDH string modified by adding the 
supersymmetric Liouville sector is analysed in the even
target space dimensions $d=2,4,6,8$.
The spectrum generating algebra for this model
is constructed
and a general version of the
no-ghost theorem is proven.  A counterpart of the GSO
projection leads to a family of tachyon free
unitary free string theories. One of these models
is equivalent to
the non-critical Rammond-Neveu-Schwarz spinning string
truncated in the Neveu-Schwarz sector
to the tachyon free eigenspace of the fermion parity
operator.
\end{abstract} 
\thispagestyle{empty}
\end{titlepage}

\section{Introduction}

The relevance of the (super-)Liouville theory for a proper
description of quantum string in non-critical dimensions 
was first pointed out in Polyakov's celebrated papers on 
conformal anomaly in bosonic \cite{polyakov81a} and fermionc
\cite{polyakov81b} string models. This observation  inspired
extensive studies of the quantum Liouville
\cite{liouville}
and the super-Liouville theory
\cite{superliouville}.
In spite of significant progress in
$D\leq 1$ non-critical strings \cite{ddk} and
superstrings \cite{superddk} and
the subsequent development of
the corresponding Liouville \cite{lcft} and
the super-Liouville \cite{superlcft} conformal field theories
the question whether the famous $D=1$ barrier can be overcome
remains open.

There are basically two different approaches to this problem.
One of them developed by
Gervais and collaborators \cite{gervais}
has already brought many promising results
but technicalities involved make it
difficult to go beyond topological models \cite{cremmer94}.
The second approach recently advocated by Polyakov 
\cite{polyakov97}
identifies the Liouville degree of freedom with 
an extra fifth curved dimension. This idea received unexpected 
support from the recently discovered relation between the
fundamental IIB superstring in $AdS_5\times S^5$ background
and the supersymmetric Yang-Mills theory in four dimensions
\cite{maldacena}. It is believed that this
approach may lead to a plausible scenario for non-critical strings
but there are still many difficult open problems \cite{polyakov98}.

Whatever an ultimate understanding of the Liouville dynamics
would be it seems reasonable to analyse less ambitious
and more elementary questions about the role of the
Liouville degrees of freedom in the quantum mechanics of free
strings. The first attempt in this direction was
made long time ago by Marnelius \cite{marnelius83}.
He considered the standard
string model modified by adding the Liouville 
sector. Assuming some general features of the quantum
Liouville dynamics he was able to show \cite{marnelius86} that
the non-zero Liouville
modes can be identified with the longitudinal Brower excitations
of the non-critical Nambu-Goto string \cite{brower72}.

This modification of bosonic string has been recently reconsidered
under the
assumption that the bulk and boundary cosmological 
constants vanish \cite{hasiewicz96}.
This apparently drastic simplification leads however to
a nontrivial free string model exhibiting many
interesting features \cite{hasiewicz96}.
In particular the constraint eliminating the Liouville
zero mode appears as a consistency
condition for the variational principle. This excludes
for instance the interpretation
of the Liouville sector in terms of an extra target
space dimension.

In the covariant quantization 
the Liouville sector is described by
a free 2-dimensional conformal field
theory of the Fegin-Fuchs type \cite{dotsenko84}.
The corresponding no-ghost theorem \cite{hasiewicz96}
admits a family of new non-critical free bosonic quantum strings 
called massive strings for the properties of their spectra -
all states except the tachyonic ground state are massive. One
of them with the largest possible space of null
states is called the {\it critical massive
string}\footnote{The model is critical with
respect to the structure of null states rather than the 
dimension $D$ of the target space which may vary in the
range $1<D<25$.}. It is completely equivalent
to the non-critical Nambu-Goto string \cite{brower72}.
In this special case  one can also develop
the light-cone formulation and
calculate the particle content of the model \cite{daszkiewicz98}.

The problem addressed in the present paper is the covariant
operator quantization of the fermionic counterpart of open massive
string and the general no-ghost theorem for this model.
The paper is organised as follows. In Section 2 the
classical model of massive string is introduced. In Section 3
we present the covariant quantization of the model.
In Section 4 the space of physical states is explicitly
constructed in terms of DDF operators. This is our main result.
It is used in Section 5 to prove
a general version of the no-ghost theorem.
In Section 6 we introduce a non-critical counterpart of the GSO
projection and construct a family of tachyon free
unitary fermionic string models.

\section{Classical theory}

One can introduce the
fermionic massive string as a world-sheet
$N=1$ supersymmetric extension of the bosonic model.
It is defined by the spinning string covariant action 
\cite{deser76}
supplemented by the supersymmetric Liouville action
\cite{martinec83,chaudhuri87,phong88}
with vanishing cosmological terms. In the superspace 
notation \cite{howe} it takes the form:
\begin{eqnarray}
S[E,\Phi,X] &=&
-{\textstyle{\alpha \over 2\pi}} \int\limits_M
\!d^2z\,d^2\theta E D_\alpha X_{\mu} D^\alpha 
X^{\mu}
\label{action} \\
& & -{\textstyle{\beta \over 2\pi}} \int\limits_M
\!d^2z\,d^2\theta E \left( D_\alpha \Phi D^\alpha 
\Phi - 2iS_E\Phi \right)
\;\;\;.\nonumber
\end{eqnarray}
where
\vspace{-\abovedisplayskip}
$$
X^\mu(x,\theta) = x^\mu(x) + i \overline{\theta}\psi^\mu
+{\textstyle{i\over 2}} \overline{\theta}\theta D^\mu\;\;\;,\;\;\;
\Phi(x,\theta) \;=\; \varphi(x) + i \overline{\theta}\psi^L
+{\textstyle{i\over 2}} \overline{\theta}\theta D^L\;\;\;,
$$
are the embedding and the Liouville 2-dim real scalar 
superfields, respectively. 
For  the 2-dim supergravity sector we use
Howe's notations and conventions \cite{howe}
($E_M^A$ denotes the supervierbein and $S_E$ is the corresponding 
curvature scalar superfield).
In all three sectors we assume  unique supersymmetric
extensions \cite{divecchia82,martinec83}
of the boundary conditions of the bosonic massive string
\cite{hasiewicz96}.
 
The action is invariant under superdiffeomorphisms and 
local Lorentz transformations. Due to the absence of 
cosmological terms it is also invariant with respect to a
special class of superconformal transformations
$$
E_M^a \longrightarrow e^\Sigma E_M^a\;\;\;\;,\;\;\;\;
E_M^\alpha \longrightarrow e^{\Sigma \over 2}E_M^\alpha
-{\textstyle{i\over 2}} e^{{\Sigma \over 2}}E_M^a
(\gamma_a)^{\alpha \beta} D_\beta \Sigma
 \;\;\;,
$$
with scaling superfields $\Sigma$ satisfying the equation
$
D_\alpha D^\alpha \Sigma = 0.
$
Due to this extra symmetry the supergravity sector can be 
completely gauged away. In
the flat superconformal gauge 
$E_M^A = \widehat{E}_M^A$ 
$$
\widehat{E}_m^a = \delta_m^a\;\;;\;\;
\widehat{E}_m^\alpha\;=\;0\;\;;
\;\;\widehat{E}_\mu^a\;=\;i\theta^\lambda(\gamma^a)_{\lambda \mu} 
\;\;;\;\;\widehat{E}_\mu^\alpha \;=\;0\;\;,
$$
the model is given by the following system of equations 
\begin{eqnarray*}
D_\alpha D^\alpha X^\mu \;=\;
D_\alpha D^\alpha \Phi &=& 0\;\;\;\;,\nonumber\\
\alpha(\gamma^b\gamma^a)_\alpha^{\;\beta}
D_\beta X^\mu \partial_b X_\mu +
\beta (\gamma^b\gamma^a)_\alpha^{\;\beta}
\left(D_\beta \Phi \partial_b \Phi - 2 D_\beta \partial_b \Phi \right) 
&=&0\;\;\;\;
\label{energymomentum}\\
\int\limits_{\partial M_i} ds\,d^2\theta\;
\theta^\alpha n_a\gamma^a D_\alpha\Phi &=&0\;\;\;\;,
\label{zeromode}
\end{eqnarray*}
where $\partial M_i$ denotes the "initial" world-sheet boundary.
The origin of the last constraint is essentially the same as in 
the bosonic massive string \cite{hasiewicz96}.
Proceeding to the components and eliminating the auxiliary fields
$D^\mu, D^L$ one has
\begin{eqnarray*}
\partial^a\partial_a x^\mu &=&\gamma^a\partial_a \psi^\mu\;=
\;\partial^a\partial_a \varphi \;=\;\gamma^a\partial_a 
\psi^L\;=\;0\;\;\;,\nonumber\\
J^b_\alpha&=&T_{ab}\;=\;
\int\limits_{\partial M_i} ds \, n^a\partial_a \varphi
\;=\;0\;\;\;,
\end{eqnarray*}
where the supersymmetry current $J^a_\alpha$ and the energy 
momentum tensor $T_{ab}$ are given by
\begin{eqnarray*}
J^a_\alpha &\equiv&
{\textstyle{\alpha\over \pi}} \partial_b x^\mu
(\gamma^b\gamma^a\psi_\mu)_\alpha +
{\textstyle{\beta \over \pi}}
\left( \partial_b \varphi
(\gamma^b\gamma^a\psi^L)_\alpha -
\eta^{ab}\partial_b{\psi^L}_\alpha \right)\;\;\;,\\
T_{ab} &\equiv& {\textstyle{\alpha\over\pi}}
\left(
\partial_ax^{\mu} \partial_bx_{\mu}
- {\textstyle{1\over 2}} \eta_{ab} \partial^c x^{\mu} \partial_c
x_{\mu} +{\textstyle{i\over 4}}
\overline{\psi}^\mu(\gamma_a\partial_b
+ \gamma_b\partial_a )\psi_\mu -
{\textstyle{i\over 4}} \eta_{ab}
\overline{\psi}^\mu \gamma^c\partial_c\psi_\mu
\right)\nonumber \\
&+ & {\textstyle{\beta\over\pi}} \left( \partial_a\varphi \partial_b\varphi
-{\textstyle{1\over 2}}\eta_{ab} \partial^c\varphi \partial_c\varphi
-2\partial_a\partial_b\varphi
+{\textstyle{i\over 4}} \overline{\psi}^L(\gamma_a\partial_b
+ \gamma_b\partial_a )\psi^L - {\textstyle{i\over 4}} \eta_{ab}
\overline{\psi}^L \gamma^c\partial_c\psi^L
\right)\;\;\;.
\end{eqnarray*}
In the flat superconformal 
gauge the supersymmetric boundary conditions 
\cite{divecchia82,martinec83} take the form
$$
\begin{array}{cllcll}
x'^\mu(\tau,0)
&=\;x'^\mu(\tau,\pi)\;=\;0\;
&,\;\;\;
&\varphi'(\tau,0)&=\;\varphi'(\tau,\pi)\;=0\;
&,\\[6pt]
\psi^\mu_+(\tau,0)
&=\;\psi^\mu_-(\tau,0)\;
&,\;\;\;
&\psi^L_+(\tau,0)&=\;\psi^L_-(\tau,0)\;&,\\[6pt]
\psi^\mu_+(\tau,\pi)&=\;(-1)^\epsilon\psi^\mu_-(\tau,\pi)\;
&,\;\;\;
&\psi^L_+(\tau,\pi)&=\;(-1)^\epsilon\psi^L_-(\tau,\pi)\;&,
\end{array}
$$
where $\epsilon =0$ corresponds to the Rammond, and $\epsilon =1$ to
the Neveu-Schwarz sector;
$\psi_\pm$ denote components of 2-dim spinor in the
basis of $\gamma^0\gamma^1$ eigenvectors.

\noindent Introducing holomorphic variables
$$
\begin{array}{rllrll}
 \dot{x}^\mu \pm x'^\mu  &=\;
{\textstyle{1\over \sqrt{\alpha}}}
\displaystyle\sum\limits_{ k \in Z\!\!\!Z}
a^\mu_k {\rm e}^{-ik(\tau \pm \sigma)}
\;&,\;
&\dot{\varphi} \pm \varphi'  &=\;
{\textstyle{1\over \sqrt{\beta}}}
\displaystyle\sum\limits_{k \in Z\!\!\!Z}
 c^\mu_k {\rm e}^{-ik(\tau \pm \sigma)}
\;&,\\[14pt]
\psi^\mu_\pm  &=\;
{\textstyle {1\over \sqrt{\alpha}}}  \!\!\!
\displaystyle\sum\limits_{r \in Z\!\!\!Z + {\epsilon \over 2} }
\!\!\! b^\mu_r {\rm e}^{-ir(\tau \pm \sigma)}
\;&,\;&
\psi^L_\pm  &=\;
{\textstyle{1\over\sqrt{\beta}}}\!\!\!
\displaystyle\sum\limits_{r \in Z\!\!\!Z + {\epsilon \over 2}}
\!\!\!d^\mu_r {\rm e}^{-ir(\tau \pm \sigma)}
\;&,\\[2pt]
q^\mu_0&=\; {\textstyle{\sqrt{\alpha}\over \pi}}
\displaystyle\int\limits_0^\pi d\sigma \; x^\mu (\sigma,0)
\;&,\;&
q^L_0 &=\; {\textstyle{\sqrt{\beta}\over \pi}}
\displaystyle\int\limits_0^\pi d\sigma \; \varphi (\sigma,0)
\;&,
\end{array}
$$
satisfying the canonical graded Poisson bracket relations
$$
\begin{array}{cllcll}
\{ a^\mu_m, a^\nu_n \} &=\; im \eta^{\mu\nu}\delta_{m,-n}
\;&,\;\;\;&
\{ c_m, c_n \} &=\; im \delta_{m,-n}
\;&,\\[6pt]
\{ a^\mu_0, q^\nu_o \} &=\;  \eta^{\mu\nu}
\;&,\;\;\;&
\{ c_0, q^L_0 \} &=\; 1
\;&,\\[6pt]
\{ b^\mu_m, b^\nu_n  \}&=\;
i\eta^{\mu\nu} \delta_{m,-n}\;&,\;\;\;&
\{ d_m, d_n  \}&=\;
i \delta_{m,-n}\;&,
\end{array}
$$
one can rewrite the constraints in the following standard
form
\begin{eqnarray}\label{constraints}
  L_m  & = & {\textstyle{1\over 2}}\!\! \sum\limits_{n \in Z\!\!\!Z}
  a_{-n}\cdot a_{n+m} \;+\;
  (1-\epsilon) {\textstyle{d+1\over 16}}\delta_{m,0}\; + \;
{\textstyle{1\over 2}}
 \sum\limits_{r \in Z\!\!\!Z + {\epsilon\over 2} }
   r \,b_{-r} \cdot b_{r+m} \nonumber\\
& + &
{\textstyle{1\over 2}} \sum\limits_{n \in Z\!\!\!Z}
 c_{-n}c_{n+m}\; +\; 2i\sqrt{\beta}m c_m \;+ \;2\beta\delta_{m0}\; +\;
{\textstyle{1\over 2}} \sum\limits_{r \in Z\!\!\!Z + {\epsilon \over 2} }
 r\, d_{-r} d_{r+m}\;\;\;, \\
G_r   & = &
 \sum\limits_{n \in Z\!\!\!Z}
a_{-n}\cdot b_{n+r}\; +\;
 \sum\limits_{n \in Z\!\!\!Z}
c_{-n}d_{n+r}\; + \;4i\sqrt{\beta} r d_r \;\;\;,\nonumber\\
c_0 &=&0\;\;\;.
\end{eqnarray}
The Poincare generators are represented by
\begin{eqnarray}
P^\mu &=& \sqrt{\alpha} a^\mu_0\;\;\;,\nonumber\\
M^{\mu\nu}& = &  x^\mu_0 P^\nu - x^\nu_0 P^\mu -
i\sum\limits_{n>0}
{\textstyle{1\over n}} \left( a^\mu_{-n} a^\nu_{n} - a^\nu_{-n} a^\mu_{n}
\right)\;\;\;\label{poincare}\\
&&- i\sum\limits_{r>0}
\left( b^\mu_{-r} b^\nu_{r} - b^\nu_{-r} b^\mu_{r}
\right) -  \epsilon i b^\mu_0 b^\nu_0 \;\;\;,  \nonumber
\end{eqnarray}
where $x^\mu_0={1\over \sqrt{\alpha}}q^\mu_0$.

\section{Covariant quantization}

Following standard   prescriptions of covariant quantization 
we start with  the algebra of canonical commutation and anticommutation 
relations
\begin{equation}
\label{ccr}
\begin{array}{cllcll}
{[a^{\mu}_m,a^{\nu}_n]}
&=\; m \eta^{\mu\nu} \delta_{m,-n}\;\;\;
&,\;\;\;
&[c_m,c_n]
&=\; m \delta_{m,-n}\;\;\;&,\\[6pt]
{[a_0^{\mu},q_0^{\nu}]}
& =\; - i \eta^{\mu\nu}\;\;\;
&,\;\;\;
&[c_0,q_0^L]
& = \;-i\;\;\;&,\\[6pt]
\{b^\mu_r,b^\nu_s \}
&=\; \eta^{\mu\nu} \delta_{r,-s}\;\;\;
&,\;\;\;
&\{d_r,d_s \}
&=\; \delta_{r,-s}\;\;\;&, 
\end{array}
\end{equation}
supplemented by  the conjugation properties
\begin{equation}
\label{conjugation}
\begin{array}{lllllllll}
\left(a^{\mu}_0\right)^+\!\!\!&=\; a^{\mu}_0\!\!\!&,\;\;
\left(q_0^{\nu}\right)^+\!\!\!&=\;q_0^{\nu}\!\!\!&,\;\;
\left(c_0\right)^+\!\!\!&=\; c_0\!\!\!&,\;\;
\left(q_0^L\right)^+\!\!\!&=\;q_0^L\!\!\!&,\\[6pt]
\left( a^{\mu}_m \right)^+\!\!\!& =\; a^{\mu}_{-m}\!\!\!&,\;\;
\left( b^\mu_r \right)^+\!\!\!&=\; b^\mu_{-r}\!\!\!&,\;\;
\left( c_m \right)^+\!\!\!& =\; c_{-m}\!\!\!&,\;\;
\left( d_r \right)^+\!\!\!& =\; d_{-r}\!\!\!&,
\end{array}
\end{equation}
where $m\in Z\!\!\!Z;r \in Z\!\!\!Z + {\textstyle{\epsilon\over 2}}$.
Let us denote by  $F_\epsilon(p,p^L)$
the Fock space generated by the algebra of non-zero modes
out of the unique vacuum state $\Omega_\epsilon(p,p^L)$ satisfying
$$
\begin{array}{clll}
a^{\mu}_m \Omega_\epsilon(p,p^L) &=\;
c_m \Omega_\epsilon(p,p^L)& = \;0\;&,\;\;\;m>0\;\;\;,\\[6pt]
b^{\mu}_r \Omega_\epsilon(p,p^L) &=\;
d_r \Omega_\epsilon(p,p^L)& =\; 0\;&,\;\;\;r>0\;\;\;,\\[6pt]
P_{\mu} \Omega_\epsilon(p,p^L) & =\;
p_{\mu} \Omega_\epsilon(p,p^L) &&,\\[6pt]
P_L \Omega_\epsilon(p,p^L) &=\;
p_L \Omega_\epsilon(p,p^L) &&.
\end{array}
$$
The space of states is a direct sum of the pseudo-Hilbert spaces
$H_\epsilon (p,p^L)$ along $(d+1)$-dimensional spectrum of
the momentum operators $P^\mu,P^L\equiv\sqrt{\beta}c_0$
$$
H_\epsilon = \int d^dp\, dp^L \, H_\epsilon(p,p^L)\;\;\;.
$$
In the Neveu-Schwarz sector ($\epsilon = 1$)
$$
H_1(p,p^L) = F_1(p,p^L)\;\;\;.
$$
In the Rammond sector ($\epsilon = 0$) the fermionic zero modes
$b^\mu_0,d_0$
form the real Clifford algebra ${\cal C}(d,1)$ corresponding to the
metric of $(d,1)$ signature.
If one requires a well defined fermion parity
operator  the zero mode sector of $H_0(p,p^L)$ must carry an
irreducible representation of the real Clifford algebra
${\cal C}(d+1,1)$.

For the sake of simplicity we restrict ourselves to the even
dimensions $d=2,4,6,8$ of the target space.
As we shall see in Sect.5. higher dimensions
are excluded by the no-ghost theorem.
Let $D(d+2)$ be the space of an  irreducible
representation of the complex extension
${\cal C}^C(d+2) = {\cal C}(d+1,1)\otimes C\!\!\!\!I$ of
the real Clifford algebra ${\cal C}(d+1,1)$.  Note that for even
$d$ the algebra  ${\cal C}^C(d+2)$ is simple and there is only one
such representation of complex dimension $2^{{d+2\over 2}}$.
$D(d+2)$ regarded as a representation
of ${\cal C}(d+1,1)$ is irreducible only
for $d=4$ \cite{coquereaux}.  For $d=2,6,8$ the complex representation
$D(d+2)$ decomposes into
a direct sum of two equivalent real irreducible representations  of
${\cal C}(d+1,1)$
$$
D(d+2) = S(d+1,1) \oplus S(d+1,1)\;\;\;,
$$
where $S(d+1,1)$ 
are characterised by appropriate Majorana conditions
\cite{coquereaux}.

Strictly speaking the rules of covariant quantization
require an irreducible representation of the real
Clifford algebra ${\cal C}(d+1,1)$. We shall however admit
the complex representation $D(d+2)$ which leads to the following
structure of $H_0(p,p^L)$ 
$$
H_0(p,p^L) = F_0(p,p^L)\otimes D(d+2) \;\;\;,
$$
and provides a unified formulation for all cases $d=2,4,6,8$.
Other reasons for this choice will be discussed in Sect.5.

Let us denote by
${\tilde a}^\mu_n,{\tilde c}_n,{\tilde b}^\mu_r,{\tilde d}_r$
the operators on $F_0(p,p^L)$ representing  non-zero bosonic and
fermionic modes.
Using gamma matrices of the $D(d+2)$ representation normalised by
$-(\Gamma^0)^2=(\Gamma^1)^2=
\dots=(\Gamma^{d-1})^2=(\Gamma^L)^2=
(\Gamma^F)^2=1$,
one can construct a representation of
the algebra (\ref{ccr}) on $H_0(p,p^L)$ as follows
\begin{equation}
\label{modes}
\begin{array}{rcllrcll}
a^\mu_n &=& {\tilde a}^\mu_n\otimes 1\;&,\;\;\;&
c_n&=& {\tilde c}_n \otimes 1\;&,\;\; n\neq 0\;,\\
b^\mu_r &=&  {\tilde b}^\mu_r\otimes \Gamma^F\;&,\;\;\;&
d_r&=& {\tilde d}_r \otimes\Gamma^F\;&,\;\; r\neq 0\;,\\
b^\mu_0 &=& 1\otimes {\textstyle {1\over \sqrt{2}}}\Gamma^\mu  \;&,\;\;\;&
d_0 &=& 1\otimes{\textstyle {1\over \sqrt{2}}}\Gamma^L \;&,\;\;
\end{array}
\end{equation}
The fermion parity operator on $H_0(p,p^L)$ can be defined by
$$
(-1)^F =  (-1)^{\tilde{F}} \otimes \Gamma^F
$$
where  $\tilde{F} = \sum_{r>0}\tilde{b}_{-r} \cdot \tilde{b}_{r}
+ \sum_{r>0}\tilde{d}_{-r}\tilde{d}_{r}$ is the fermion number
operator on $F_0(p,p^L)$.

The hermitian scalar products in the spaces $F_0(p,p^L)$,
$F_1(p,p^L)$, and
$D(d+2)$ are  determined
by the relations (\ref{ccr}) and the conjugation properties
(\ref{conjugation}).

The constraint operators $L_m, G_r$ on $H_\epsilon$ are represented
by normally ordered counterparts of the classical expressions
(\ref{constraints}). For the bosonic and fermionic zero modes we
assume the symmentric and the antisymmetric ordering, respectively.
In both sectors the quantum constraints algebra takes the form
\begin{eqnarray*}
   \left[ L_m, L_n \right]   & = & (m-n)L_{m+n} +
                       {\textstyle\frac{1}{8}}(d+1+32\beta)(m^3-m)\delta_{m,-n}
                             \;\;\;,   \\
   \left[ L_m, G_r \right]   & = & ({\textstyle\frac{m}{2}} - r)G_{m+r}
   \;\;\;,\\
   \left\{ G_r, G_s \right\} & = & 2L_{r+s}
                + {\textstyle{1\over 2}}(d+1+32\beta)(r^2-
{\textstyle{1\over 4}})\delta_{r,-s}\;\;\;.
\end{eqnarray*}
The subspace ${\cal H}_\epsilon^{\rm ph}\subset
H_\epsilon$ of physical states is defined by the conditions
$$
\begin{array}{rllll}
  ( L_n  - a_\epsilon\delta_{n,0} )\Psi&=\;0 &,\;&
   n \geq 0 &,\\[6pt]
  ( G_r + m_0 g_0\delta_{r,0})\Psi&=\;0 &,\;&
   r \geq {\epsilon\over 2} &,\\[6pt]
   (P^L - p^L) \Psi&=\;0 &,&&
\end{array}
$$
where $g_0=1\otimes\Gamma^F$, and
$a_\epsilon,m_\epsilon, p^L \in {I\!\!R}$, are
regarded as parameters of
the quantum model. They are related by the condition
$$
a_\epsilon = -m_\epsilon^2 +
(1-\epsilon){\textstyle{1\over 16}}(d+1)
 +2\beta\;\;\;.
$$
The "mass" operator $g_0$
has been introduced in order to
obtain the most general setting
in which the conditions involving $L_0$, and $G_0$ are consistent
with the algebra relation  $G_0^2=L_0 -
{1\over 16}(d+1+32\beta)$.

The normal ordering does not alter
the classical expressions for the Poincare generators (\ref{poincare})
and their algebra
is represented on ${\cal H}_\epsilon^{\rm ph}$ without anomaly.

\section{DDF construction}

The DDF construction \cite{ddf} for the Neveu-Schwarz model was
developed long time ago by Schwarz \cite{schwarz72},
and by Brower and Friedman \cite{brower73}. 
In this section we shall
extend this technique to the fermionic massive string.
For this purpose it is convenient to introduce
the subspace  ${\cal H}_\epsilon\subset H_\epsilon$
of on-shell states $\Psi$ satisfying the on-mass-shell
conditions\footnote{The
on-mass-shell conditions are not
essential for the construction of DDF operators. We impose them at
the beginning in order to  simplify our presentation.}
\begin{equation}
\label{on-shell}
 (L_0 - a_\epsilon)\Psi=
\left({\textstyle{1\over 2\alpha}} P^\mu P_\mu +
{\textstyle{1\over 2\beta}} ( P^L)^2  + R +m_\epsilon^2\right)
\Psi \;=\;0\;\;\;,\;\;\;P^L\Psi\;=\;p^L\Psi
\end{equation}
where
$$
  R = \sum_{m>0}\left( a_{-m}a_m + c_{-m}c_m\right) +
 \sum_{r>0} r\left( b_{-r}b_r + d_{-r}d_r \right)\;\;\;,
$$
is the level operator.
Since  $L_0$ commutes with the Poincare generators
${\cal H}_\epsilon$ carries a representation of the
Poincare algebra. The decomposition of ${\cal H}_\epsilon$ 
with respect to the mass coincides with
the level structure
\begin{equation}
\label{deco}
{\cal H}_\epsilon
= \bigoplus_{N \geq 0} {\cal H}^N_\epsilon \;\;\;;\;\;\;
R\;{\cal H}^N_\epsilon = N{\cal H}^N_\epsilon \;\;\;.
\end{equation}
Each subspace ${\cal H}^N_\epsilon$ is a direct integral of
finite dimensional subspaces
${\cal H}^N_\epsilon(p)$ with fixed on-shell momentum $p$
\begin{equation}
\label{shell}
   {\cal H}^N_\epsilon = \int_{S^N_\epsilon} d\mu^N(p)\;
   {\cal H}^N_\epsilon(p)\;\;\;,
\end{equation}
where $S^N_\epsilon $ denotes  the mass shell at level $N$
determined by the condition (\ref{on-shell}), and
$d\mu^N(p)$ is the Lorentz invariant measure on $S^N_\epsilon $.

Following \cite{schwarz72,brower73} we introduce the DDF fields
$$
\begin{array}{rcllrclll}
\widetilde{X}^{\mu}(\theta) &=&
q_0^{\mu} + a_0^{\mu}\theta +
\displaystyle\sum\limits_{m\ne0}\frac{i}{m}a_m^{\mu}{\rm e}^{-im\theta}
&,&
\Phi(\theta) &=&
q_0^L + c_0^L\theta +
\displaystyle\sum\limits_{m\ne0}\frac{i}{m}c_m{\rm e}^{-im\theta}
&,\\[6pt]
\widetilde{P}^{\mu}(\theta) &=& \mbox{$\widetilde{X}^{\mu}$}'(\theta)
&,&
\Pi(\theta) &=& \Phi'(\theta)
&,\\[6pt]
\widetilde{\Psi}^{\mu}(\theta) &=&
\displaystyle\sum\limits_{r \in Z\!\!\!Z + {\epsilon\over 2} }
\psi_r^{\mu}{\rm e}^{-ir\theta}
&,&
\Psi_L(\theta) &=&
\displaystyle\sum\limits_{r \in Z\!\!\!Z + {\epsilon\over 2} }
\psi^L_r{\rm e}^{-ir\theta}&,
\end{array}
$$
where prime denotes differentiation with respect to $\theta$.

 Let us fix a light-cone frame
$\{e_{\pm}, e_1,\dots,e_{d-2}\}$
in $d$-dimensional Minkowski space  
normalised by
$e^2_\pm =0$, $e_+\cdot e_- =-1$, and $e_i\cdot e_j =\delta_{ij}$ for
$i,j=1,\dots,d-2$. We shall use the following  notation for the
light-cone components of a vector $V$
$$
V_\pm = e_{\pm}\cdot V\;\;,\;\;V^i\;=\;e_i\cdot V\;\;\;.
$$

In the original construction the domain of DDF operators
is restricted to states with momenta
satisfying ${1\over \sqrt{\alpha}} p_+ \in Z\!\!\!Z$.
This restriction can be overcome introducing
slightly modified DDF fields
$$
\begin{array}{rcllrcll}
X_{\pm}(\theta)&=&
\left({\sqrt{\alpha}\over p_+}\right)^{\pm 1}\widetilde{X}_\pm(\theta)
 &,&
X^i(\theta)&=&\widetilde{X}^i{\mu}(\theta)&,\\[6pt]
P_{\pm}(\theta)&=& \left({\sqrt{\alpha}\over p_+}\right)^{\pm 1}
\widetilde{P}_{\pm}(\theta)
&,&
 P^i(\theta)&=&\widetilde{P}^i{\mu}(\theta)&,\\[6pt]
\Psi_{\pm}(\theta)&=& \left({\sqrt{\alpha}\over p_+}\right)^{\pm 1}
\widetilde{\Psi}_{\pm}(\theta)
 &,&
 \Psi^i(\theta)&=&\widetilde{\Psi}^i(\theta)
 &.
\end{array}
$$
The rescaling above leaves unchanged all the (anti)commutation
relations relevant for calculating the algebra of DDF operators.
Let us note that
a single light-cone frame is sufficient for a global
DDF parameterisation of ${\cal H}_\epsilon$.
Indeed the momenta with $p_+=0$ for which
the construction gets singular form
a zero measure subset of each mass shell $S^N_\epsilon$ and
can be neglected.

The DDF operators for the "super-matter" sector are exactly the
same as in the Neveu-Schwarz model \cite{schwarz72,brower73}.
In our notation they read
\begin{eqnarray*}
A^i_m &=&  {\textstyle{1\over 2\pi}}\int\limits_0^{2\pi} \!\!d\theta\,
:(P^i - m\Psi^i\Psi_+)
\,{\rm e}^{imX_+}:\;\;\;,\\
B^i_r &=&  {\textstyle{1\over 2\pi}}\int\limits_0^{2\pi} \!\!d\theta\,
:(\Psi^iP_+^{{1\over2}} - P^i\Psi_+ P_+^{-{1\over 2}}
-{\textstyle{i\over 2}} \Psi^i\Psi_+\Psi_+' )
\,{\rm e}^{irX_+}:\;\;\;,\\
A^-_m &=&  {\textstyle{1\over 2\pi}}\int\limits_0^{2\pi} \!\!d\theta\,
:(P_- -m\Psi_-\Psi_+ )
\,{\rm e}^{imX_+}:\;\;\;\\*
&-&{\textstyle{1\over 2\pi}}\int\limits_0^{2\pi} \!\!d\theta\,
:{\textstyle{i\over 2}}(m P_+^{-1}P_+' + m^2\Psi_+\Psi_+'P_+^{-1})
\,{\rm e}^{imX_+}:\;\;\;,\\
B^-_r &=&  {\textstyle{1\over 2\pi}}\int\limits_0^{2\pi} \!\!d\theta\,
:(\Psi_-P_+^{{1\over2}} - P_-\Psi_+ P_+^{-{1\over 2}}
-{\textstyle{i\over 2}} \Psi_-\Psi_+\Psi_+' )
\,{\rm e}^{irX_+}:\\*
&+&{\textstyle{1\over 2\pi}}\int\limits_0^{2\pi} \!\!d\theta\,
:\left[{\textstyle{1\over 8}}(\Psi_+P_+^{-1})'P_+^{-{3\over 2}} P_+'
- {\textstyle{5\over 4}}(\Psi_+P_+^{-1})'P_+^{1\over 2}  \right.     \\*
&&\left.
\hspace{70pt} +\; {\textstyle{\epsilon\over 16}} \Psi_+P_+^{-{3\over 2}}
-{\textstyle{i\over 8}} \Psi_+\Psi_+'\Psi_+''P_+^{-{7\over 2}} \right]
\,{\rm e}^{irX_+}:\;\;\;.
\end{eqnarray*}
Note that the antisymmetric ordering of fermionic zero modes
is responsible for the sector dependent term in the
expression for $B^-_r$.

In order to find appropriate DDF operators for
the super-Liouville excitations one can start with the "naive"
operators 
\begin{eqnarray*}
C^0_m &=&  {\textstyle{1\over 2\pi}}\int\limits_0^{2\pi} \!\!d\theta\,
:(\Pi -m\Psi_L\Psi_+ )
\,{\rm e}^{imX_+}:\;\;\;\\
D^0_r &=&  {\textstyle{1\over 2\pi}}\int\limits_0^{2\pi} \!\!d\theta\,
:(\Psi_LP_+^{{1\over2}} - \Pi\Psi_+ P_+^{-{1\over 2}}
-{\textstyle{i\over 2}} \Psi_L\Psi_+\Psi_+' )
\,{\rm e}^{irX_+}: \;\;\;,
\end{eqnarray*}
and calculate their (anti)commutators with the fermionic constraints
\begin{eqnarray*}
\left[G_r,C^0_m\right] &=&-
{\textstyle{1\over 2\pi}}\int\limits_0^{2\pi} \!\!d\theta\,
r{\rm e}^{ir\theta}:4i\sqrt{\beta}
\Psi_+ \,{\rm e}^{imX_+}:\;\;\;,\\
\left\{G_r,D_s^0\right\} &=&
{\textstyle{1\over 2\pi}}\int\limits_0^{2\pi} \!\!d\theta\,
r{\rm e}^{ir\theta}:4i\sqrt{\beta}
\,(P_+^{{1\over2}} - {\textstyle{i\over 2}}
\Psi_+\Psi_+' P_+^{-{3\over 2}}) \,{\rm e}^{isX_+}:\;\;\;.
\end{eqnarray*}
The corrections required
can be deduced  by
comparing the expressions above  with the corresponding 
calculations  for  the "naive" longitudinal DDF operators
\cite{schwarz72,brower73}.
In the case of the bosonic Liouville operator $C_m = C^0_m +
\delta C_m$ the correction  
$$
\delta C_m = {\textstyle{1\over 2\pi}}\int\limits_0^{2\pi} \!\!d\theta\,
:2\sqrt{\beta}\,(P_+^{-1}P_+' + m\Psi_+\Psi_+'P_+^{-1})
\,{\rm e}^{imX_+}:\;\;\;.
$$
is proportional to
the bosonic longitudinal correction \cite{schwarz72,brower73}.
The correction for the fermionic Liouville operator
$D_r = D^0_r + \delta D_r$ is not that obvious.
It can be however identified as a part of
the fermionic longitudinal correction 
derived in \cite{schwarz72,brower73}
$$
\delta D_r = 
-{\textstyle{1\over 2\pi}}\int\limits_0^{2\pi} \!\!d\theta\,
:4\sqrt{\beta}\,(\Psi_+P_+^{-1})'P_+^{1\over 2}\,{\rm e}^{irX_+}:\;\;\;.
$$

The transverse and the Liouville DDF operators satisfy
the canonical (anti)commutation relations
\begin{equation}
\label{ddftr}
\begin{array}{cllcll}
{[A^i_m,A^j_n]}
&=\; m \delta^{ij} \delta_{m,-n}\;\;\;
&,\;\;\;
&[C_m,C_n]
&=\; m \delta_{m,-n}\;\;\;&,\\[6pt]
\{B^i_r,B^j_s \}
&=\; \delta{ij} \delta_{r,-s}\;\;\;
&,\;\;\;
&\{D_r,D_s \}
&=\; \delta_{r,-s}\;\;\;&,
\end{array}
\end{equation}
and all cross (anti)commutators between these two families vanish.

The longitudinal DDF operators  $A^-_m, B^-_r$
do not commute with other DDF operators.
Following   \cite{brower72}
one can diagonalize the algebra
introducing "shifted" longitudinal operators
\begin{eqnarray}
\label{shifted}
A^L_m &=& A^-_m - {\cal L}_m + {\textstyle{1\over 2}}\delta_{m,0}
\;\;\;,\\
B^L_m &=& B^-_r - {\cal G}_r\;\;\;,\nonumber
\end{eqnarray}
where
\begin{eqnarray*}
{\cal L}_m  & = & {\textstyle{1\over 2}}\!\! \sum\limits_{n \in Z\!\!\!Z}  
A^i_{-n} A^i_{n+m} \;+\;
  (1-\epsilon) {\textstyle{d-1\over 16}}\delta_{m,0}\; + \;
{\textstyle{1\over 2}}
 \sum\limits_{r \in Z\!\!\!Z + {\epsilon\over 2} }
   r\, B^i_{-r}  B^i_{r+m} \nonumber\\
& + &
{\textstyle{1\over 2}} \sum\limits_{n \in Z\!\!\!Z}
 C_{-n}C_{n+m}\; +\; 2i\sqrt{\beta}m C_m \;+ \;2\beta\delta_{m0}\; +\;
{\textstyle{1\over 2}} \sum\limits_{r \in Z\!\!\!Z + {\epsilon \over 2} }
 r\,D_{-r} D_{r+m}\;\;\;, \\
{\cal G}_r   & = &
{\textstyle{1\over 2}} \sum\limits_{n \in Z\!\!\!Z}
A^i_{-n} B^i_{n+r}\; +\;
{\textstyle{1\over 2}} \sum\limits_{n \in Z\!\!\!Z}
C_{-n}D_{n+r}\; + \;4i\sqrt{\beta} r d_r \;\;\;.
\end{eqnarray*}
The new longitudinal operators
commute with the transverse and the Liouville DDF operators
and form an $N=1$ superconformal algebra
with the central charge $\hat{c}=9-d -32\beta$
\begin{eqnarray}
   \left[ A^L_m, A^L_n \right]   & = & (m-n)A^L_{m+n} +
            {\textstyle\frac{1}{8}}(9-d-32\beta)(m^3-m)\delta_{m,-n}
                             \;\;\;, \nonumber  \\
   \label{ddflo}
   \left[ A^L_m, B^L_r \right]   & = &
   ({\textstyle\frac{m}{2}} - r)B^L_{m+r}
   \;\;\;,\\
   \left\{ B^L_r, B^L_s \right\} & = & 2A^L_{r+s}
                + {\textstyle{1\over 2}}(9-d-32\beta)(r^2-
{\textstyle{1\over 4}})\delta_{r,-s}\;\;\;.\nonumber
\end{eqnarray}
By construction all the DDF operators ${\cal D}_a$,
$2a\in Z\!\!\!Z$
satisfy the following basic relations
$$
\begin{array}{rcllll}
[ \,L_m,{\cal D}_a ] & =& [\,G_r, {\cal D}_a \} \;=\;0&,
&m \in Z\!\!\!Z\;\;,\;\;r\in Z\!\!\!Z
+{\textstyle {\epsilon \over 2}}\;\;&,\\[6pt]
[ \, R\,,{\cal D}_a ] & =& -\,a\, {\cal D}_a &,& &\\[6pt]
{\cal D}_a\, \Omega_\epsilon(p) & =&0 &,&
a>0&,
\end{array}
$$
where the on-shell vacua $\Omega_\epsilon(p)\in {\cal H}^0_\epsilon(p)$
are defined as $0$-level states with the on-shell momentum $p\in S^0_\epsilon$.

In the Neveu-Schwarz sector the space ${\cal H}^0_1(p)$
is 1-dimensional and the unique on-shell vacuum
$\Omega_1(p)$ is physical.
In the Rammond sector the space ${\cal H}^0_0(p)$
carries a finite dimensional representation of the Clifford algebra.
The subspace ${\cal H}^0_0(p)^{\rm ph}\subset {\cal H}^0_0(p)$
of physical vacuua is determined by the condition
\begin{equation}
\label{gzero}
G_0\, \Omega_0(p) = {\textstyle{1\over 2}}
\left( {p_\mu\over \sqrt{\alpha}} b_0^\mu +
{p^L\over \sqrt{\beta}} d_0 + m_0 g_0
\right)\,\Omega_0(p)\;=\;0\;\;\;.
\end{equation}
We introduce a common for both sectors notation
$\left\{ \Omega^J_\epsilon(p) \right\}_{J \in {\cal J}_\epsilon}$
for a basis in the space ${\cal H}^0_\epsilon(p)^{\rm ph}$ of
physical vacuua with a momentum $p$.

Let us consider monomials $\vartheta[{\cal D}]$ of the DDF
operators ${\cal D}_a$ with negative indices $a<0$, and with
some fixed ordering. With the restriction $a<0$ the ordering matters
only for the longitudinal DDF operators where one can choose
for instance the standard ordering of the superconformal Verma module
construction. We denote by
$\left\{ \vartheta^{(N)}_I[{\cal D}] \right\}_{I\in
{\cal I}^{(N)}}$ the collection of all ordered normalised monomials of the DDF
operators with negative indices and with a fixed level $N$
$$
\left[\,R\,,\,\vartheta^{(N)}_I[{\cal D}]\right] = N\,
\vartheta^{(N)}_I[{\cal D}]\;\;\;.
$$
With the above notation one gets
\medskip

\noindent{\bf Lemma } {\it
For any $N>0$ and $p\in S^N_\epsilon$, $p_+ \neq 0$ the DDF states
$$
\vartheta^{(N)}_I[{\cal D}] \,
\Omega^J_\epsilon\left(p + {\textstyle{\alpha N\over p_+}} e_+\right)
\;\;\;\;\;\;\;\;\;J\in {\cal J}_\epsilon\;\;,\;\;
I\in {\cal I}^{(N)}\;\;,
$$
form a basis in the subspace
${\cal H}^N_\epsilon(p)^{\rm ph}\subset {\cal H}^N_\epsilon(p)$
of all physical states with the momentum $p$.}\medskip

The lemma can be proven using Brower's ideas \cite{brower72}
in exactly the same way as in the case of the Neveu-Schwarz
model \cite{brower73}.

\section{No-ghost theorem}

For any $p_+\neq 0$, and $\overline{p}= p^ie_i$
let us consider the subspace
${\cal H}_\epsilon(p)^{\rm ph}$
of all states generated by the DDF operators out of
the physical vacua $\Omega_\epsilon(p)\in
{\cal H}^0_\epsilon(p)^{\rm ph}$ with the momentum
$$
p = -{\textstyle{\alpha\over p_+}}
({\textstyle {1\over 2\beta}}p_L^2 + m^2_\epsilon )e_+
-p_+e_- + \overline{p}\;\;\;.
$$
It follows from the algebra of DDF operators
(\ref{ddftr},\ref{ddflo}) that
${\cal H}_\epsilon(p)^{\rm ph}$
has the structure of the tensor product
\begin{equation}
\label{tensor}
{\cal H}_\epsilon(p)^{\rm ph} =
{\cal F}^{\rm tr} \otimes{\cal F}^L \otimes
{\cal V}_{\epsilon}({\hat{c},h_\epsilon})\otimes
{\cal H}^0_\epsilon(p)^{\rm ph} \;\;\;,
\end{equation}
where ${\cal F}^{\rm tr}, {\cal F}^L$ denote Fock spaces
generated by the algebra
of the transverse and the super-Liouville DDF operators
(\ref{ddftr}), and ${\cal V}_\epsilon({\hat{c},h_\epsilon})$
is the superconformal Verma module
generated by the Rammond-Neveu-Schwarz superconformal algebra
of the "shifted" longitudinal DDF operators with
the  central charge $\hat{c}= 9-d- 32 \beta$ (\ref{ddflo}).
The highest weight
$h_\epsilon$ of ${\cal V}_\epsilon(\hat{c},h_\epsilon)$
is determined by the
structure of the "shifted" longitudinal operator $A^L_0$ (\ref{shifted}),
and the on mass shell condition (\ref{on-shell})
$$
h_\epsilon =
{\textstyle{1\over 2} }
-(1-\epsilon){\textstyle{1\over 16}}(d-1)
- 2\beta + m^2_\epsilon\;\;.
$$

Using the light-cone parameterisation of the mass shells
$S^N_\epsilon $ and the lemma of Sect.4 one can represent
the subspace ${\cal H}_\epsilon^{\rm ph}$
of all physical states in the following form
\begin{equation}
\label{decompo}
{\cal H}_\epsilon^{\rm ph}
=  \int {dp_{+}\over |p_+|} d^{d-2}\overline{p}\;
{\cal H}_\epsilon(p)^{\rm ph}\;\;\;.
\end{equation}

The metric structure
of each ${\cal H}_\epsilon(p)^{\rm ph}$
is completely determined by the algebra of DDF operators
(\ref{ddftr},\ref{ddflo}) along with their conjugation properties
with respect to the metric structure of the original pseudo-Hilbert
space $H_\epsilon$:
$$
\begin{array}{cllcllcll}
(A^i_m )^\dagger
&=\; A^i_{-m}\;\;
&,\;\;
&(A^L_m)^\dagger
&=\; A^L_{-m}\;\;
&,\;\;
&(C_m )^\dagger
&=\; C_{-m}\;\;
&,\;\;\;m\in Z\!\!\!Z\;,\\[6pt]
( B^i_r )^\dagger
&=\; B^i_{-r}\;\;
&,\;\;
&( B^L_r )^\dagger
&=\; B^L_{-r}\;\;
&,\;\;
&(D_r)^\dagger
&=\;D_{-r}\;\;
&,\;\;\;r \in Z\!\!\!Z + {\textstyle{\epsilon\over 2}}\;.
\end{array}
$$
The Fock space component  ${\cal F}^{\rm tr} \otimes{\cal F}^L$
of ${\cal H}_\epsilon(p)^{\rm ph}$ (\ref{tensor})
is the same in both sectors and carries a positive definite metric.
The metric structure of the superconformal Verma module
${\cal V}_\epsilon({\hat{c},h_\epsilon})$ had been extensively
studied in the context of unitary highest weight representations
of the Rammond-Neveu-Schwarz superconformal algebra.
The necessary  conditions for the absence of negative norm states
in ${\cal V}_\epsilon({\hat{c},h_\epsilon})$ were derived by Friedan, Qiu, and 
Shenker
\cite{friedan84a,friedan84b}. It was farther proven by
Goddard, Kent, and Olive \cite{goddard86}
that these conditions are also sufficient.

In the Neveu-Schwarz sector
the subspace ${\cal H}^0_1(p)^{\rm ph}$ is 1-dimensional and
the decomposition (\ref{tensor}) simplifies
$$
{\cal H}_1(p)^{\rm ph} =
{\cal F}^{\rm tr} \otimes{\cal F}^L \otimes
{\cal V}_1({\hat{c},h_1})\;\;\;.
$$
Then the results concerning the metric structure of
${\cal V}_1(\hat{c},h_1)$
\cite{friedan84a,friedan84b,goddard86}
yield the following
\bigskip

\noindent{\bf No-Ghost Theorem -- Neveu-Schwarz sector}

{\it The space of physical states in the Neveu-Schwarz sector of
the fermionic massive string is ghost free if and only if one
of the following two conditions is satisfied

\noindent 1. continuous series
$$
0< \beta \leq {8-d\over 32} \;\;\;\;,\;\;\;\;
m^2_1 \geq 2\beta - {1\over 2}\;\;\;,
$$
2. discrete series
$$
\beta = \beta_m = {8-d\over 32} + {1\over 4m(m+2)}
\;\;\;,\;\;\;
m^2_1 = m^2_{p,q} = -{d\over 16} + {[(m+2)p - mq]^2 \over 8m(m+2)}
\;\;\;,
$$
where $m,p,q$ are integers satisfying $2\leq m$, $1\leq p< m$,
$1\leq q <m +2$, $p-q$ even.
}\bigskip\bigskip

In the Rammond sector the structure of ${\cal H}_0(p)^{\rm ph}$
is more complicated
$$
{\cal H}_0(p)^{\rm ph} =
{\cal F}^{\rm tr} \otimes{\cal F}^L \otimes
{\cal V}_0({\hat{c},h_0})\otimes
{\cal H}^0_0(p)^{\rm ph} \;\;\;.
$$
Let us assume that  the
necessary and sufficient conditions for the non-negative
metric structure of the superconformal Verma module of the Rammond
superalgebra
\cite{friedan84a,friedan84b,goddard86}
are satisfied. Then the properties of the metric structure
of ${\cal H}_0^{\rm ph}$ are entirely determined by the properties of the
metric structures of the subspaces of physical
vacuua ${\cal H}^0_0(p)^{\rm ph}$.

In order to find an explicit form
of the scalar product in  ${\cal H}^0_0(p)^{\rm ph}$
let us first analyse  hermitian scalar products on
the irreducible representations $D(d+2)$ of the complex
Clifford algebras ${\cal C}^C(d+2)$. There
exists one and only one hermitian positive definite
scalar product $(\;\;,\;\;)$ on $D(d+2)$ such that
all gamma matrices are isometries.
We define the parity operator
$$
\theta =\left\{ \begin{array}{ll}
                i\Gamma^0\Gamma^1\dots\Gamma^{d-1}\Gamma^L\Gamma^F &
                {\rm for}\; d=2,6\\
                \Gamma^0\Gamma^1\dots\Gamma^{d-1}\Gamma^L\Gamma^F &
                {\rm for}\; d=4,8
               \end{array} \right. \;\;\;.
$$
Note that $\theta$ is hermitian
with respect to $(\;\;,\;\;)$ and $\theta^2 =1$.
The rules of covariant quantization require
a hermitian scalar product on $D(d+2)$ for which
$\Gamma^\mu, \Gamma^L, \Gamma^F$ are all hermitian.
A unique  product with
these properties is given by
$$
\langle \xi, \zeta \rangle =
(\xi, \Gamma^0 \theta \,\zeta ) \;\;\;, \;\;\xi,\zeta \in D(d+2)\;\;.
$$
The 0-level physical states
$\psi\in {\cal H}_0^{0\,{\rm ph}}$ satisfy
$G_0$ constraint (\ref{gzero}). In the
position representation it takes the form of the Dirac equation
\begin{equation}
\label{dirac}
\left( -{i\over \sqrt{2\alpha}} \Gamma^\mu \partial_\mu +
{p^L\over \sqrt{2\beta}} \Gamma^L + m_0\Gamma^F
\right) \psi = 0\;\;\;.
\end{equation}
For any two solutions $\psi(x),\phi(x)$ the vector current
$$
j^\mu(x) = \langle \psi(x),   \Gamma^\mu \phi(x) \rangle
$$
is conserved, and can be used for constructing
a Lorentz  invariant  scalar product
on ${\cal H}_0^{0\,{\rm ph}}$
\begin{eqnarray}
\label{product}
\langle \psi ,\phi \rangle& =&
\int dx^1\dots dx^{d-1}\;
(\psi(x),\theta
\phi(x) )\\
& =&
\int dx^-d^{d-2}\overline{x}\;
(\psi(x),\Gamma^+\Gamma^0 \theta
\phi(x) )\;\;\;.
\nonumber
\end{eqnarray}
Proceeding to the momentum
representation and taking into account the Lorentz
invariant measure in (\ref{decompo}) one
gets the scalar product
in ${\cal H}^0_0(p)^{\rm ph}$
$$
\langle \psi(p),\phi(p) \rangle_{\rm l.c.}  =
(\psi(p),\Gamma^+\Gamma^0 \theta
\phi(p) )\;\;\;.
$$
Since it is neutral so is
the metric structure of
${\cal H}_0(p)^{\rm ph}$,
and in consequence the metric structure of the space 
${\cal H}^{\rm ph}_0$ of all physical states.
This is not necessarily a disaster for the unitarity of the model.
What saves the day is a Lorentz invariant decomposition of
${\cal H}^{\rm ph}_0$ into a direct sum of two orthogonal
components
\begin{equation}
\label{dec}
{\cal H}^{\rm ph}_0=
{\cal H}^{\rm ph}_{0\;+} \oplus {\cal H}^{\rm ph}_{0\;-}
\end{equation}
with a positive and a "negative" definite scalar products,
respectively. In order to construct such
decomposition  let us consider the extension $\Theta_0 = \theta \otimes 1$
of $\theta$ from $D(d+2)$ to the whole
pseudo-Hilbert space $H_0$ of covariant quantization.
Using the explicit realisation (\ref{modes}) one can easily verify
that $\Theta_0$ has all the properties of the fermion parity
operator. This in particular implies that $\Theta_0$ commutes
with the Lorentz generators, and anticommutes
with the Dirac operator (\ref{dirac}).
Then it follows from the representation (\ref{product})
that the eigenspaces ${\cal H}_{0\;\pm}^{\rm ph}$
corresponding to $\pm 1$ eigenvalues of
$\Theta_0$ yield the decomposition required.

The unitarity problem can be solved by imposing
the superselection rule related to the
decomposition (\ref{dec}).
This is possible if and only if an operator $\Theta_0$
with the required properties can be constructed which
is always the case if we start with the complex representation
$D(d+2)$. 
Note that for $d<9$ and within the covariant
quantization based on the irreducible 
representations $S(d+1,1)$ of the Clifford algebra 
${\cal C}(d+1,1)$
the operator $\Theta_0$  exists only for $d=4$ and $d=8$
\cite{coquereaux}. This provides another justification for
our choice made in Sect.3.

Taking into account the restrictions
for possible central
charges, and highest weights of the superconformal
Verma module ${\cal V}_0(\hat{c},h_0)$
of the Rammond superconformal algebra one gets the following
\bigskip

\noindent{\bf No-Ghost Theorem -- Rammond sector}

{\it The space ${\cal H}^{\rm ph}_0$ of
physical states in the Rammond sector
andmits a unique
Lorentz invariant scalar product.
This product is neutral.
${\cal H}^{\rm ph}_0$ decomposes
into an orthogonal  direct sum
$$
{\cal H}^{\rm ph}_0=
{\cal H}^{\rm ph}_{0\;+} \oplus {\cal H}^{\rm ph}_{0\;-}
$$
of the eigenspaces of the parity operator $\Theta_0$.

The eigenspaces  ${\cal H}^{\rm ph}_{0\;\pm}$ are
ghost free if and only if one
of the following two conditions is satisfied

\noindent 1. continuous series
$$
0< \beta \leq {8-d\over 32} \;\;\;\;,\;\;\;\;
m^2_0 \geq 2\beta - {8-d\over 16}\;\;\;,
$$
2. discrete series
$$
\beta = \beta_m = {8-d\over 32} + {1\over 4m(m+2)}
\;\;\;,\;\;\;
m^2_0 = m^2_{p,q} = {[(m+2)p - mq]^2 \over 8m(m+2)}
\;\;\;,
$$
where $m,p,q$ are integers satisfying $2\leq m $, $1\leq p< m$,
$1\leq q <m +2$, $p-q$ odd. 
}\bigskip\bigskip

In all our considerations
the zero mode of the
Liouville momentum $p^L$ has been regarded as a
free parameter of the quantum theory.
The no-ghost theorems
do not impose any restriction on $p^L$. Since there is
no ordering ambiguity in the
constraint $P^L=0$ we shall assume $p^L=0$. In this case
the parameters $m_0,m_1$ have the interpretation of
the physical masses of the corresponding ground states.

\section{Conclusions}

The results of the previous
section yield  a complete classification of all admissible
Hilbert spaces ${\cal H}^{\rm ph}_\epsilon(\beta,m_\epsilon)$
of both sectors of the massive fermionic string
in terms of the parameters $\beta$, and $m_\epsilon$.
For $\beta$ from the discrete series
the total Hilbert space is given by
$$
{\cal H}^{\rm ph}(m) =
\bigoplus_{p-q\;{\rm odd}}{\cal H}^{\rm ph}_{0}(\beta_m,m_{p,q})
\bigoplus_{p-q\;{\rm even}}{\cal H}^{\rm ph}_{1}(\beta_m,m_{p,q})
\;\;\;,
$$
where the sum runs over
all $1\leq p<m,\; 1\leq q <m+2$, and $m=2,3,\dots\;$.
The central charge $\hat{c}$ of the superconformal algebra of the
"shifted" longitudinal DDF operators
$$
\hat{c}= \hat{c}_m \;=\; 1 - {8\over m(m+2)}\;\;\;,
$$
"measures" the number of
longitudinal physical degrees of freedom.
In particular for $m=2$, one gets $\hat{c}_2=0$, and all states
containing longitudinal excitations are null.
The quantum theory contains one "superfunctional"
physical degree of freedom less than the classical one.
For large $m$ the central charge $\hat{c}_m$ approaches 1.
The limiting case $\hat{c}=\hat{c}_\infty =1$ corresponds
to $\beta ={1\over 32} (8-d)$ -- the upper bound of
the continuous series.
We define
$$
{\cal H}^{\rm ph}(\infty) =
{\cal H}^{\rm ph}_{0}({\textstyle{1\over 32}} (8-d),0)
\oplus
{\cal H}^{\rm ph}_{1}({\textstyle{1\over 32}} (8-d),
-{\textstyle{1\over 16}}d)
\;\;\;.
$$
In this model the space of physical states is largest possible.
Both the classical, and the quantum theories contain
$d$ "superfunctional" physical degrees of freedom.

For $\beta$ in the range $0<\beta<{1\over 32}(8-d)$,
one has $1<\hat{c}$, and
the structure of the physical degrees of freedom of the quantum
theory is essentially the same as in the model
described by ${\cal H}^{\rm ph}(\infty)$. One could in principle
define a continuous family of models but
without any extra physical assumption the ground state
masses $m_0,m_1$  are undetermined free parameters of such construction.

Models  from the
family $\left\{{\cal H}^{\rm ph}(m)\right\}_{m=2}^\infty$
are not quite satisfactory.
The spectrum of each Neveu-Schwarz sector contains
tachyon, while the metric in the Rammond sector
is always neutral.
As we have seen in the previous section the second problem
can be overcome by introducing superselection rule
related to the operator $\Theta_0$.
One can try to extend this
rule to the whole Hilbert space.
Following the scheme of the GSO projection
in the critical string \cite{gliozzi77} we
decompose the space ${\cal H}^{\rm ph}_1(m)$ of
the Neveu-Schwarz sector
into the direct sum
$$
{\cal H}^{\rm ph}_1(m)=
{\cal H}^{\rm ph}_{1\;+}(m)\oplus {\cal H}^{\rm ph}_{1\;-}(m)
$$
of the $\pm 1$ eigenspaces of the
fermion parity operator $\Theta_1$
$$
\Theta_1= (-1)^{F+1}\;\;\;,\;\;\;F =
\sum\limits_{r>0}
(b_{-r}\cdot b_r + d_{-r}d_r )\;\;\;.
$$
With this definition of $\Theta_1$
the eigenspace ${\cal H}^{\rm ph}_{1\;+}(m)$ corresponding to $+1$
eigenvalue does not contain the tachyonic ground state.
We introduce a non-critical counterpart of the GSO projection
as the projection on the $+1$ eigenspace
$$
{\cal H}^{\rm ph}_+(m)=
{\cal H}^{\rm ph}_{0\;+}(m)\oplus {\cal H}^{\rm ph}_{1\;+}(m)
$$
of the fermion parity operator $\Theta = \Theta_0 \oplus \Theta_1$.
It yields  a family of tachyon free unitary
non-critical fermionic strings.
In contrast the original GSO projection 
\cite{gliozzi77} it does not lead to a supersymmetric spectrum.
Note that there are no massles states in models with odd $m$.

The model corresponding to beginning of the
discrete series $\beta = \beta_2 = {9-d\over 32}$
is especially interesting. In this case
the subspace of null physical
states is largest possible.
For this reason it will be called
the {\it critical  (fermionic) massive string}.
The space of "true" physical states
${\cal H}^{\rm tr}(2)$ is given by the quotient
$$
{\cal H}^{\rm tr}(2)
= {{\cal H}^{\rm ph}(2)\over
\{ {\rm null\; states}\}}\;\;\;.
$$
As in the case of the bosonic critical massive string
\cite{daszkiewicz98} one can show that the subspace
${\cal H}^{\rm RNS}=
{\cal H}^{\rm RNS}_0\oplus {\cal H}^{\rm RNS}_1 \subset
{\cal H}^{\rm ph}(2)$
generated by the transverse $A^i,B^i$,
and the longitudinal $A^-,B^-$, DDF operators
is a good gauge slice for the quotient map
${\cal H}^{\rm ph}(2)\rightarrow {\cal H}^{\rm tr}(2)$.
In consequence the even (with respect to
$\Theta$) subspace ${\cal H}^{\rm RNS}_+=
{\cal H}^{\rm RNS}_{0\;+}\oplus {\cal H}^{\rm RNS}_{1\;+}$
yields a 1-1 parameterisation of the quotient
space  ${\cal H}^{\rm tr}_+(2)$  of the GSO projected model.

The superalgebra of the transverse
and the longitudinal DDF operators
of the critical massive string
is by construction isomorphic
with the whole DDF algebra of the non-critical
Rammond-Neveu-Schwarz string \cite{schwarz72,brower73}.
This implies that in the Neveu-Schwarz sector
the subspace ${\cal H}^{\rm RNS}_1$
 is isomorphic
with the tachyon free eigenspace of the fermion parity
operator in the space of physical states of
the non-critical RNS string.
In the Rammond sector each eigenspace 
${\cal H}^{\rm RNS}_{0\;\pm}$ of $\Theta_0$
carries the same complex representation $D(d)$ of the
real Clifford algebra ${\cal C}(d-1,1)$, and is therefore
isomorphic with the whole space of
physical states in the Rammond sector of the
non-critical RNS string.

One can easily verify that the subspace
${\cal H}^{\rm RNS}_+$ is stable
with respect to the Poincare transformations, and carries
a representation which is isomorphic with
the representation of the Poincare group
in the non-critical RNS string.  It follows that
the GSO projected fermionic critical
massive string is  equivalent with the
non-critical RNS string
truncated in the Neveu-Schwarz sector to the
tachyon free eigenspace of the fermion parity
operator.

We have shown that the covariant quantization of the
free fermionic massive string in the even dimensions
$d=2,4,6,8$ leads to a family of new tachyon free
unitary fermionic strings. One of these models -
the critical fermionic massive string - is closely related
to the non-critical RNS string.

There are at least two interesting open problems
before one may try
to attack the question of the interacting theory.
First of all it would be desirable to develope
the light-cone formulation of the critical massive
string and to calculate its spin content.
In view of the equivalence  discussed above
it would provide
the particle spectrum of the non-critical RNS string.
The second problem is to analyse
the superconformal field theory structure of the
fermionic massive string. This would in particular clarify
the status of the non-critical GSO projection
proposed in this paper.
\bigskip

\noindent{\bf Acknowledgements}

The authors would like to thank Marcin Daszkiewicz for many stimulating 
discussions.
This work is supported by the Polish Committee of Scientific Research
(Grant Nr. PB 1337/PO3/97/12).

\end{document}